# Effect of time of day on reward circuitry. A discussion of Byrne et al. 2017.

Adam Steel*[1,2], Cibu Thomas[1], Chris I Baker[1]

1 - Laboratory of Brain and Cognition, National Institute of Mental Health, National Institutes of Health, Bethesda, MD, 20814

2 - Wellcome Centre for Integrative Neuroimaging, FMRIB, John Radcliffe Hospital, University of Oxford, Headington, Oxford, Oxfordshire, OX3 9DU, UK

*Corresponding author:

Adam Steel

National Institutes of Health

10 Center Drive

Bethesda, MD 20814

(202) 640 9340

adam.steel@nih.gov

Time of day has broad effects on human and animal physiology that impact all levels of organism function, from gene expression (Reppert and Weaver, 2002) to endocrine release (Czeisler et al., 1999). Given the ubiquitous impact of time of day on physiology, it is unsurprising that many behaviors (Evans et al., 2011) and psychological states also exhibit circadian-related fluctuation (Golder and Macy, 2011). Despite our growing understanding of micro- and macroscopic fluctuations that occur over the course of a day, we are only beginning to grasp the effect of these fluctuations on human behavior, and how these effects manifest at the systems-level in the human brain (Muto et al. 2016). Improving our knowledge of this mechanism could lead to novel treatments for mood disorders, which have been linked to disturbances in circadian rhythms and sleep (McClung, 2013). One recent study published in the *Journal of Neuroscience* addressed a new aspect of this topic, namely how time of day modulates the neural response to reward in humans (Byrne et al., 2017). Here, we briefly review this study and discuss both methodological and inferential concerns as well as avenues for future research.

Strong evidence suggests that subcortical structures, including dopaminergic nuclei in the midbrain, may be modulated by time of day (Muto et al. 2016). Indeed, prior to the study by Byrne and colleagues, two other studies had examined the impact of time of day on reward salience in humans using functional magnetic resonance imaging (fMRI). Activity in the striatum was greater in the evening than the morning for monetary rewards (Hasler et al., 2014) and greater in the morning than evening for rewarding food stimuli (Masterson et al., 2016). However, these studies were limited by testing only two time points, which makes them unable to detect any non-linear relationships. Byrne and colleagues sought to provide a fuller description of the variation in the reward response over the course of the day by recruiting 16 young male participants for three MRI sessions, at 10:00, 14:00, and 19:00 h, with all sessions completed within 24 h. During the fMRI sessions, participants performed a gambling reward task similar to the task used by the Human Connectome Project (Van Essen et al. 2013). Briefly, participants guessed a card value and received monetary reward for correct guesses. Unbeknownst to the participants, the values of the cards were predetermined such that reward predominated in certain blocks and loss predominated in others.

Based on their *a priori* hypotheses, the authors focused on multiple reward-related regions (including bilateral putamen, caudate, medial prefrontal cortex, and anterior cingulate cortex). Using small volume correction for these regions of interest, the authors report a significant effect of time of day on the BOLD response during reward blocks in the left putamen only (see Figure 1 from Byrne et al. 2017). Subsequently, a follow-up analysis focused on the voxels discovered in this primary analysis. Based on a model that considered the three time points, the authors report a 'periodic waveform,' with a decreased response to reward (compared to baseline) at 14:00 h compared to 10:00 and 19:00 h. Finally, a whole-brain analyses using a liberal cluster threshold correction revealed clusters in the left insula and middle frontal gyrus, although it is unclear how the putamen results compare to this whole-brain threshold.

Based on these results, the authors conclude that BOLD activation in response to reward varies over the course of the waking day, with reward evoking a lower response at 14:00 h compared to 10:00 and 19:00 h. The authors discuss this result in the context of reward prediction error (Schultz et al., 1997) and propose that reward is anticipated differentially throughout the day. The authors acknowledge that their results do not replicate the findings from prior work either in direction of effect or anatomical location (Masterson et al. 2016; Hasler et al. 2014). These discrepancies are attributed to the use of different rewarding stimuli compared to one study (Masterson et al. 2016), and differences in the scan time and experimental contrasts compared to the other (Hasler et al. 2014).

Overall, Byrne and colleagues examine a timely topic with potentially important results. However, issues with the analytical methods and experimental design significantly limit the interpretability and generalizability of their findings. We focus on four major points, beginning with methodological issues before moving on to more general considerations.

First, the statistical approach employed in the paper limits confidence in the robustness of the findings. The primary findings in the paper are based on small volume correction implemented across a suite of regions. While appropriate in some cases, small volume correction requires that the regions chosen must be not only strongly motivated, but also explicitly defined (Poldrack et al., 2017). In the present paper, the authors justify the regions chosen based on their involvement in reward processing. However, the regions being tested are not formally defined: no information is provided about their anatomical extent or number of voxels, they are not shown in the figure, and it is unclear whether small volume correction was applied to each region independently, which would require multiple comparisons correction for the number of regions being tested (Poldrack et al., 2008, 2017). The authors also report results which explicitly do not survive small volume correction (including cluster sizes of 4 and 1 voxel), which should not be interpreted. Beyond the *a priori* defined regions, the whole brain analysis reveals additional regions (prefrontal and insular cortex) where reward-evoked activity is modulated by time of day, but it is unclear whether the left putamen shows significant modulation in this whole-brain analysis. Finally, while the manuscript implies that the time of day modulation is specific to the left putamen, this is not explicitly tested. Rather, it is primarily based on the qualitative difference between the regions of interest. In order to show that the effect of time of day is specific to the putamen, a direct comparison should be made between the putamen and the other regions tested.

Second, the follow-up test used to confirm the nature of time of day effect is inherently circular. Specifically, the selection of voxels for this test is based on the presence of significant modulation by time of day determined by the first analysis. This virtually guarantees the presence of a significant effect (Kriegeskorte et al., 2009). Moreover, the motivation to test the specific quadratic model (equivalent morning and evening; lower in the afternoon) presented by the authors is not clear given that prior work found the opposite pattern of results (Hasler et al. 2014). Thus, it seems likely that this model was chosen post-hoc. To resolve these issues, the quadratic model could have been: i) tested on an independent dataset, ii) tested on the average activity within the *a priori* anatomically-defined ROIs, or iii) the current data could have been split in half to separate voxel and model selection from model testing. Given these concerns, it will be important to replicate these findings in an independent dataset.

Third, in the context of brain systems, it is imperative to understand how the physiological changes interact with neuroimaging measures. When studying time of day processes, it is important to consider that cortisol (Krieger et al., 1971; Montoya et al., 2014), caffeine (Park et al., 2014), and activities like food intake and hydration all vary systemically and influence physiological responses and brain imaging measures. Time of day is known to impact T1-weighted images and therefore measures like cortical thickness and apparent grey matter density (Nakamura et al., 2015; Trefler et al., 2016). In the functional domain, resting state connectivity and regional cerebral blood flow of the default mode network varies based on time of day (Hodkinson et al., 2014). Notably, cerebral blood flow of anterior cingulate cortex, which is involved in emotional regulation and reward processing, shows regional cerebral blood flow that is related to salivary cortisol levels (Hodkinson et al. 2014). These highly complex interactions between time of day, physiology, and measurement techniques complicate interpretation of the present study. However, relating any observed neural changes to behavior in the same participants would strengthen the significance and implications of the reported effects (Thomas and Baker, 2013; Krakauer et al., 2017). Although Byrne and colleagues discuss cyclical variation in mood reported in other studies (Murray et al. 2009; Murray et al.

2002; Boivin et al. 1997), these measures are not reported for the participants in their study. Therefore, the relevance of these findings to human cognition is unclear.

Finally, although the manuscript makes a broad claim regarding reward system function, the limited experimental conditions employed do not provide adequate evidence for the paper's general conclusion. The paper used only a single task and a single stimulus. Because of the limited context, whether these findings generalize to other rewarding stimuli and other tasks is unclear. The discrepancy between the current findings and prior work highlights this concern (Hasler et al. 2014; Masterson et al. 2016). The issue of generalizability could be resolved by utilizing multiple tasks within a single subject or by using a large variety of stimuli (Westfall et al. 2016). In addition, to make a strong assertion that the reward-related activity in putamen varies broadly over the time of day, future studies will need to demonstrate that the time of day modulation is specific to reward prediction error or saliency, as opposed to other functions like movement preparation or statistical learning.

When considering future work, several manipulations might be used to make strong assertions regarding modulation of reward system function based on time of day in a biologically relevant fashion. One possibility is to explicitly map the change in the activity over the course of a day for different types of rewards. Alternatively, sensitivity to different forms of reward learning, such as approach-avoidance or reinforcement learning, might be tracked over the course of the day. Either of these experiments would yield data that would aid interpretation, and massive online behavioral data collection could be used to obtain a robust estimate of effect sizes, which would aid the design of future studies of reward-related behavior.

In summary, the study by Byrne and colleagues contributes an interesting hypothesis to chronobiological research. However, limitations in the analytical methods and experiment design, as well as the absence of a direct link between their finding and behavior, makes it difficult to conclude that neural activation in response to reward fluctuates with time of day, since the changes in putamen activation could be epiphenomenal or due to confounding factors that were not accounted for in the study. Future work should integrate more robust behavioral data or leverage the statistical power of large datasets such as the Human Connectome project to understand better the impact of time of day on functional and structural properties of the brain.

## Acknowledgements

A.S., C.T., and C.I.B. were supported in part by the NIMH Intramural Research Program (NIH intramural program number ZIA MH002893-10).